# Probing charge scattering mechanisms in suspended graphene by varying its dielectric environment

A.K.M. Newaz[1], Yevgeniy S. Puzyrev[1], Bin Wang[1], Sokrates T. Pantelides[1], and Kirill I. Bolotin[1,*]

[1]Department of Physics and Astronomy, Vanderbilt University, Nashville, TN-37235, USA

*e-mail: kirill.bolotin@vanderbilt.edu

**Graphene with high carrier mobility $\mu$ is required both for graphene-based electronic devices and for the investigation of the fundamental properties of graphene's Dirac fermions[1]. It is largely accepted that the mobility-limiting factor in graphene is the Coulomb scattering off of charged impurities that reside either on graphene or in the underlying substrate[2]. This is true both for traditional graphene devices on $SiO_2$ substrates[2] and possibly for the recently reported high-mobility suspended[3,4] and supported[5] devices. An attractive approach to reduce such scattering is to place graphene in an environment with high static dielectric constant $\kappa$ that would effectively screen the electric field due to the impurities[6,7,8,9]. However, experiments so far report only a modest effect of high-$\kappa$ environment on mobility[10,11]. Here, we investigate the effect of the dielectric environment of graphene by studying electrical transport in multi-terminal graphene devices that are suspended in liquids with $\kappa$ ranging from 1.9 to 33. For non-polar liquids ($\kappa$<5) we observe a rapid increase of $\mu$ with $\kappa$ and report a record room-temperature mobility as large as ~60,000 cm$^2$/Vs for graphene devices in anisole ($\kappa$=4.3), while in polar liquids ($\kappa$>18) we observe a drastic drop in $\mu$. We demonstrate that non-polar liquids enhance mobility by screening charged impurities adsorbed on graphene, while charged ions in polar liquids cause the observed mobility suppression. Furthermore, using molecular dynamics simulation we establish that scattering by out-of-plane flexural phonons, a dominant scattering mechanism in suspended graphene in vacuum at room temperature[12], is suppressed by the presence of liquids. We expect that our findings may provide avenues to control and reduce carrier scattering in future graphene-based electronic devices.**

To vary the dielectric constant of graphene's environment controllably, we fabricated six large (2-4μm by 8-10μm) multiprobe graphene devices that are suspended in liquids with the dielectric constant $\kappa$ varying from ~1.9 to ~33 (Fig. 1). The liquids are non-polar solvents - hexane ($\kappa$=1.9), toluene (2.3), anisole (4.3) as well as polar liquids - isopropanol (18), ethanol (25) and methanol (33). Large leakage currents prevented us from measuring the devices in solvents with higher $\kappa$, such as water ($\kappa$=79). We expect that changes of $\kappa$ should significantly affect electrical transport in suspended devices since both sides of the graphene sheet are exposed to the high-$\kappa$ medium and since substrate-induced scattering is effectively eliminated, leaving only scattering from charged impurities on the surface of graphene. During the course of the experiments, we study the electric transport parameters – conductivity ($\sigma$), Hall carrier mobility ($\mu$) at $n$=5x10$^{11}$ cm$^{-2}$, and effective capacitance ($C_g$) using the same device suspended in different liquids at room temperature (RT) and under ambient conditions (Fig. 1). To ensure that no additional scatterers are adsorbed onto the device between the measurements, we never dried the devices during the experiment (see Methods).

Polar and non-polar liquids have a very different effect on electrical transport in suspended graphene (Fig. 2a). In non-polar liquids, we observe an increase of $\mu$ ($=\sigma/ne$ with $n$=5x10$^{11}$ cm$^{-2}$) with $\kappa$ for every measured device (Fig. 2b). In a representative device (Device #1), the mobility increases from $\mu$~29,000 cm$^2$/Vs in hexane ($\kappa$=1.9) to a large value $\mu$~45,000 cm$^2$/Vs in anisole ($\kappa$=4.3), more than twice the value for the same device on a substrate (Fig. 2a, dashed curve). In a different device (Device #2), the mobility in anisole reached ~60,000 cm$^2$/Vs. To the best of our knowledge, this is the highest reported mobility for a graphene device at room temperature and close to the highest room temperature mobility seen in any semiconducting material[13] (while higher mobility values are reported in a very recent paper[14], these values are measured at a lower temperature ~250K and at a lower carrier density). The devices are stable over days of measurements and exhibit changes in $\mu$ of <10% upon cycling multiple times through various solvents. At the same time, despite large changes in $\mu$, the minimal conductivity $\sigma_{min}$ is not affected by graphene's dielectric environment (Fig. 2a, Inset).

Qualitatively, the behavior observed in non-polar liquids is consistent with Coulomb scattering due to charged impurities/residues adsorbed on the graphene surface being the dominant scattering mechanism. Indeed, the potential due to such impurities is expected to be strongly screened in solvents with higher $\kappa$[6,7,8,9]. This makes our observation of the drop in mobility in polar solvents with $\kappa>18$, sometimes to values lower than the same device in air, especially surprising (Figs. 2a,b). We propose that the lowering of the mobility indicates an additional scattering mechanism that dominates in polar liquids – Coulomb scattering of graphene charge carriers by charged ions that are present in polar liquids.

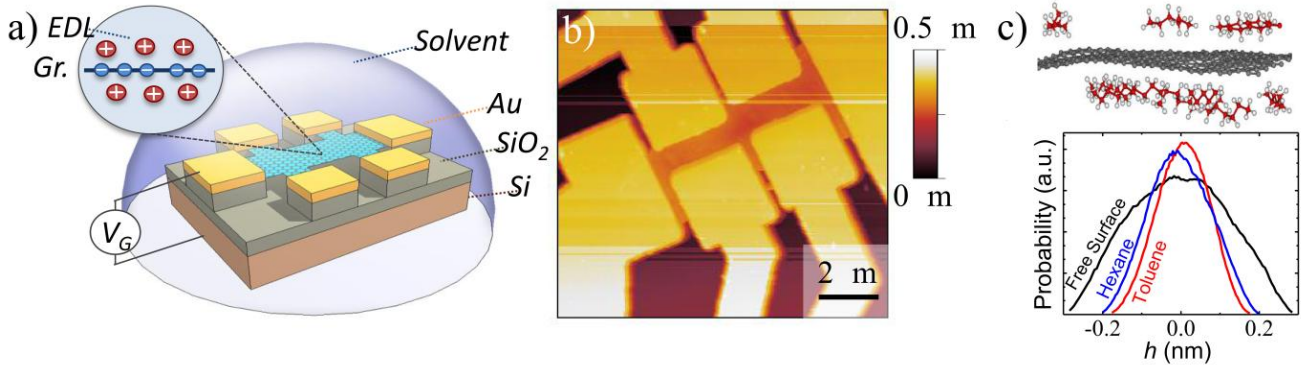

Figure 1: **Multi-terminal graphene device suspended in liquids with varying dielectric constant.** a) Device schematics: gold electrodes support a graphene sheet ~200nm above the SiO$_2$/ Si substrate; liquid surrounds the entire device. When potential is applied between the graphene and the gate electrode, an ionic electric double layer (EDL) forms next to the graphene. b) The suspended graphene device dried and imaged in air using an atomic force microscope after the completion of measurements. Imaging confirms that the device remained suspended during the course of the experiments, rather than collapsed onto the substrate. The specimen shown is smaller than the typical devices used in the experiments. c) Molecular dynamics simulations of flexural phonons (ripples) in suspended graphene in liquids and in vacuum at room temperature. Top panel: a snapshot from the simulation of graphene in hexane (not all the hexane molecules are shown). Bottom panel: heights (*h*) distribution of a graphene sheet suspended in hexane, toluene, and in vacuum. Note that the ripples are significantly suppressed in non-polar solvents.

Indeed, charged ions, likely the results of contamination with ambient water vapor or trace impurities, are always present in liquids at ambient conditions, with concentrations greater for polar than in non-polar liquids. Bulk conductivity measurements of the liquids used in our experiments allow us to estimate the molar concentration of charged ions as 30-50 mM for polar and <10 μM for non-polar solvents (see Supplementary Materials (SM)). To quantify the presence and the distribution of the ions, we examine the variation of the device capacitance $C_g$ in different liquids. While $C_g$ measured in non-polar solvents is close to the values obtained for devices on the SiO$_2$(300nm)/Si substrates, $C_g$ ~120aFμm$^{-2}$, the capacitance reaches values up to >1×10$^4$ aFμm$^{-2}$ in polar solvents, such as in ethanol (Fig. 2c). We interpret the increase of capacitance as a simple consequence of the so-called *electrolyte gating*[15]. Since the back-gate electrode is in contact with the liquid, the electrical potential of the bulk liquid acquires values that are a fraction of the back-gate voltage $V_g$ and an ionic electrical double layer (EDL) with a characteristic thickness *d* (*Debye length*[16]), forms next to the graphene (Fig. 1a). The formation of an EDL results in a strong electric field at the graphene-liquid interface, which in turn results in a large apparent back gate capacitance[16]. Using the Gouy-Chapman-Stern model, we estimate $d\sim 2\varepsilon_0 \kappa/C_g$, and obtain $d\gtrsim 140$ nm for non-polar and $d\lesssim 12$nm for polar solvents (details in SM). Therefore, in polar solvents the ions are in close proximity to the graphene and can contribute to Coulomb scattering (Fig. 2a, Inset).

To confirm the role of the charged ions in liquids limiting the mobility of graphene, we also studied the same suspended graphene device#1 in a non-polar solvent, anisole, into which we artificially introduced charged ions by adding a tetrabutylammonium tetraphenylborate (TBATPhB) salt in concentrations varying from 0 to 80 mM (Fig. 3a). As expected, the increased ionic concentration resulted in a drop of mobility (Fig. 3c) and an increase in capacitance (Fig. 3b), reflecting the increased scattering and higher electric fields due to the ions being on average closer to the graphene at a higher salt concentration.

*Discussion:* Our data feature two main trends: 1) large enhancement of mobility with $\kappa$ for devices in non-polar solvents, and 2) relatively low values of mobility for devices in polar solvents. To describe this behavior semi-quantitatively, we employ a model that includes only two sources of scattering: 1) Coulomb scattering from charged impurities with density $n_{imp}$ - likely the fabrication residues[17] - located on the graphene sheet[18], and 2) Coulomb scattering from ions in solution belonging to the EDL (we discuss possible contribution of other scattering mechanisms later). Since electric fields are absent in the bulk of the solution, the areal density of these ions $n_{ion}$ has to be equal and opposite in polarity to the carrier density $n$ in graphene. The spatial distribution of ions away from graphene is derived from the Gouy-Chapman-Stern model[16], while the average thickness of the EDL $d$ is obtained from the $C_g$ measurements. Both types of scattering mechanisms are screened by the dielectric surrounding the graphene. The scattering rate due to each mechanism is calculated using the semi-classical theory developed by Adam *et al.*[18], and finally Matthiessen's rule is used to get the total rate (details in SM).

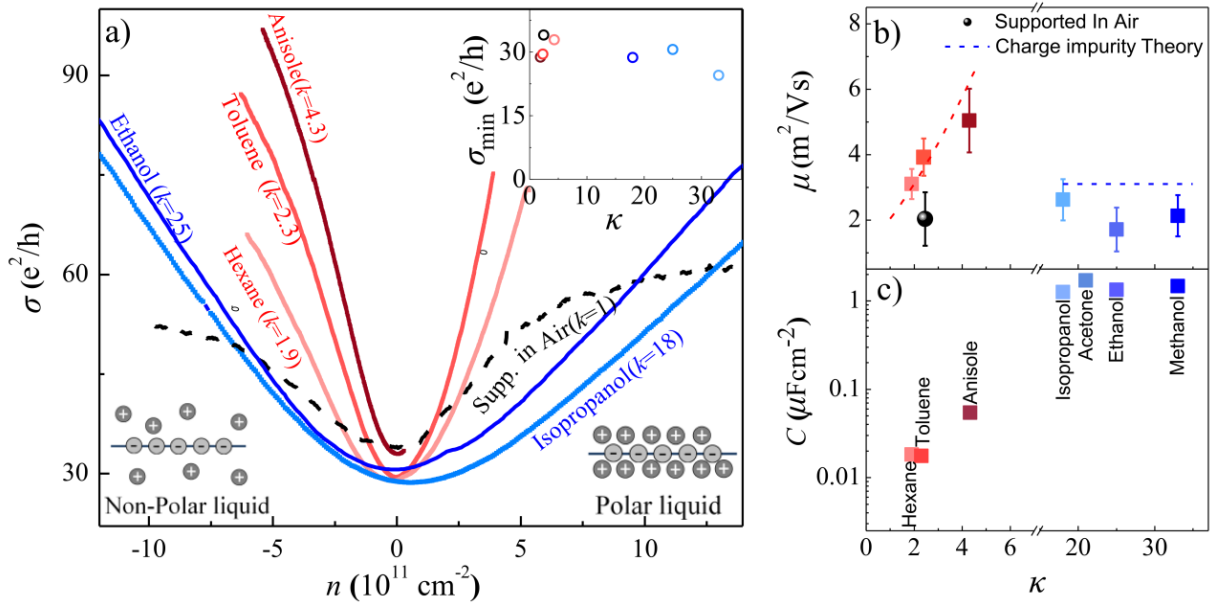

Figure 2: **Effect of the dielectric environment on the transport properties of a suspended graphene device.** *a)* The conductivity $\sigma$ as a function of carrier density $n$ for representative suspended device#1 in different liquids. For comparison, the dashed line represents $\sigma(n)$ of the same device#1 supported on SiO$_2$ and not covered by any liquid. (That data were obtained before the substrate under the device was etched away during the fabrication stage). The inset (top): the minimal conductivity $\sigma_{min}$ of the same device vs. the dielectric constant $\kappa$ of the liquid. *b)* The mobility $\mu$ at $n=5\times10^{11}$ cm$^{-2}$, averaged over 6 measured suspended graphene devices as a function of $\kappa$ of the liquid (square symbols, not every device was measured in every liquid). The black circle is the average mobility measured for 10 reference graphene devices supported on SiO$_2$ and not covered by any liquid. The red dashed line is the estimated mobility limited by Coulomb scattering from charged impurities adsorbed on graphene at a concentration $n_{imp}=1\times10^{11}$ cm$^{-2}$. The blue dashed line is a model that includes ion-induced Coulomb scattering, charged-impurity scattering with the same $n_{imp}$ and with $\kappa$ fixed at $\kappa=6$. *c)* The measured effective gate capacitance $C_G$ for the device#1 vs. $\kappa$ of the liquid surrounding it.

In non-polar solvents, where the ionic concentration is negligible (<10 μM), we assume that the Coulomb scattering from charged impurities is the only scattering mechanism at work. We fit our data treating $n_{imp}$ as the only variable parameter (Fig. 2b, red dashed line) and obtain a good fit with a realistic $n_{imp}\sim1\times10^{11}$ cm$^{-2}$. As expected, this value is lower than the values found for the supported graphene and higher than that for suspended graphene in high vacuum[18]. Remarkably, for the case of TBATPhB in anisole, where both scattering mechanism are at play, a reasonable fit is obtained using the same $n_{imp}$ and *no adjustable parameters* (dashed line in Fig. 3c). On the other hand, for polar solvents, such as ethanol, the model predicts almost complete suppression of both scattering mechanisms due to screening and a very high mobility $\mu$>30 m$^2$/Vs, much larger than observed in the experiment. We resolve this conundrum by noting that at the EDL, the bulk dielectric constant of a liquid ($\kappa$=25 in case of ethanol) can be suppressed by an order of magnitude due to the preferential orientation of polar molecules next to the metal interface[19,20]. While to our knowledge there is no detailed theory to describe $\kappa$ of the EDL, we note that the $\mu$ values measured for polar solvents are consistent with the effective $\kappa\sim6$, which is close to $\kappa$ measured for the interfacial layer of water[20] (Fig. 2b, blue dashed-line). We also note that Coulomb scattering due to the dipole moments of polar molecules[21] can be an additional scattering mechanism leading to a decrease in mobility in polar solvents.

It is instructive to consider one interesting aspect of the ion-induced carrier scattering. Since $n_{ions}=-n$, the semiclassical estimate for conductivity limited by ion and impurity Coulomb scattering[8] at first glance yields $\sigma(n)\sim n/|n_{imp}+n_{ion}|\sim n/|n_{ion}|\sim const$ for at high carrier densities ($n>>n_{imp}$). This is different from the roughly linear dependence observed in experiments (Figs. 2a,3a). However, this contradiction can be resolved by remembering that the thickness $d$ of the EDL also decreases with $n$, and hence a weaker scattering is expected at lower carrier densities. Precise modeling of the $\sigma(n)$ due to ion-induced scattering is outside the scope of this paper and awaits an appropriate theoretical model. However, the dependence $n_{ion}=-n$ means that the surface density of ions in the EDL is exactly zero when graphene is at its charge neutrality point (CNP), $n=0$. Thus, we expect that the electron transport in graphene at the CNP is unaffected by the ion-induced scattering, and that the scattering is dominated by surface-bound impurities. This prediction is consistent with the observed behavior of minimum conductivity $\sigma_{min}$ of our devices, which fluctuates <10% in the same device across the entire range of polar and non-polar solvents (Fig. 2a, Inset). Indeed, a self-consistent theory[8] predicts the variation in $\sigma_{min}$ less than 10% in the range of $\kappa$=2-33 assuming a constant $n_{imp}$. Similar nearly constant $\sigma_{min}$ was also observed in an experiment where $\kappa$ of graphene's environment was adjusted, albeit in much smaller range[10].

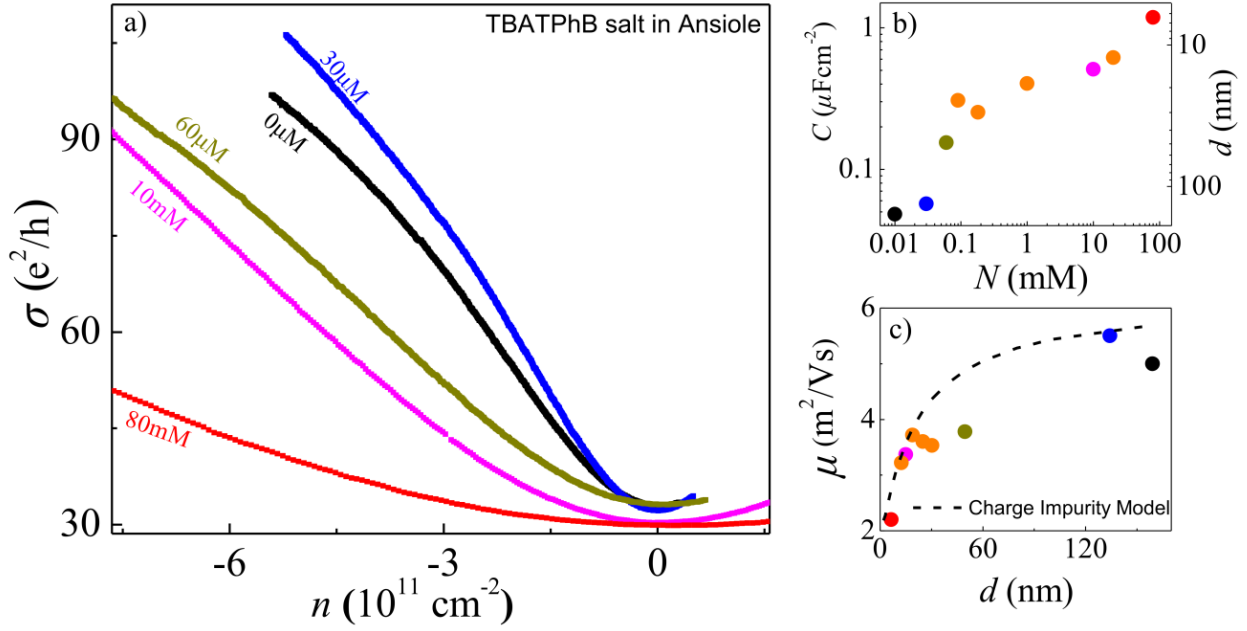

Figure 3: **Effect of the ions in the liquid on electrical transport of suspended graphene device.** The device#1 was studied in anisole to which TBATPhB salt was added in concentrations 0–80mM. a) Conductivity $\sigma(n)$ for different concentrations of the salt. b) Capacitance $C_g$ and the average thickness of the EDL $d$ estimated from $C_g$ vs. salt concentration for the same device. c) The mobility $\mu$ vs. the EDL thickness $d$. The dashed line is the expectation of the Coulomb scattering model which includes the scattering by the impurities with the concentration $n_{imp} \sim 1\times 10^{11}$ cm$^{-2}$ attached to the graphene surface and the contribution due to ions in solution.

Finally, we analyze the implicit assumption of our model that the Coulomb scattering is the dominant scattering mechanism in our devices. We note that recent experiments reported strong scattering of charge carriers in graphene by out-of-plane (flexural) phonons for suspended graphene in vacuum[12]. The values of the mobility observed here are significantly larger than the mobility limitation $\mu$<30,000 cm$^2$/Vs due to scattering on flexural phonons[12]. To resolve this seeming contradiction, we performed molecular dynamics simulations of graphene sheets suspended in either hexane, toluene or in vacuum at room temperature. We find that the interaction of molecules of the liquid with graphene suppresses the amplitude of the phonons by ~50% (Fig. 1c). We also verified computationally that this suppression is equivalent to an effective increase of the bending rigidity[22] of graphene $k$ from a free-space value ~1.3 eV in vacuum to ~3.6 eV for graphene suspended in hexane or toluene. This, in turn, translates to a mobility limitation due to phonons ($\mu_{lim}\sim$const$\times k^2$ from Ref. 12) of ~230,000 cm$^2$/Vs. Since this is significantly larger than the values of $\mu$ in our devices, we conclude that the scattering by the out-of-plane acoustic phonon is insignificant in our experiments due to the suppression of these phonons in the presence of a liquid (details in SM).

Our observations may have several important consequences. First, the demonstrated increase of mobility in non-polar liquids with high $\kappa$ may provide a viable approach towards engineering of high-mobility graphene devices that operate at room temperature. Second, we resolve the apparent contradiction between the previously reported experimental data for graphene in high-$\kappa$ environment[10,11] and the Coulomb scattering theory[6,7,8,9]. Third, the demonstrated sensitivity of electron transport in graphene to the presence of ions in solution may lead to a new paradigm of electrochemical sensors and biosensors[23]. Finally, we expect that the rich physics of ion-electron interaction encountered here may stimulate the development of a theory describing electron transport of graphene in ionic solutions. *Note:* While Chen et al. reported ultrahigh mobility in supported graphene devices covered by liquids[24], they later showed that those claims were unfounded[25].

*Methods:* The suspended graphene devices in liquid (Fig. 1a) are prepared following previous work[3]. Briefly, graphene is obtained by micromechanical exfoliation, gold/chrome electrodes are fabricated via electron beam lithography followed by metal evaporation, and the sacrificial $SiO_2$ is removed via etching in hydrofluoric acid. Crucially, the devices are never dried after etching to avoid the collapse of graphene onto the substrate due to surface tension of the drying liquid[3]. Instead, the etchant was slowly replaced by DI water and then by the high-$\kappa$ liquid under study (details in SM). A total of six suspended devices in different liquids were studied. In control experiments, we also examined ten graphene devices supported on $SiO_2$ in air and four supported devices on $SiO_2$ with liquids on top of them. The details of the electrical measurements are presented in the Supplementary Materials. These measurements are performed under ambient conditions at room temperature. We use standard four-probe measurements at low magnetic field B(0-45mT) to determine the Hall resistivity $\rho_{xy}$, the carrier density $n=B/e\rho_{xy}$, the gate capacitance $C_g=edn/dV_G$ ($V_G$ is the gate voltage), and the carrier mobility $\mu(n_{max})=1/\rho_{xx}ne$ at the maximum density $n_{max}=5\times10^{11}$ cm$^{-2}$ accessible in our devices. Unlike the suspended devices studied by others[3,4], our specimens were never current-annealed.

*Acknowledgements:* We thank Shaffique Adam for allowing us the use of his code; David Cliffel, James Dickerson, Len Feldman, and Eli Shkrob for useful discussions; Hiram Conley and Shuren Hu for technical assistance. We acknowledge support through NSF CAREER DMR-1056859 and DTRA grant HDTRA1-10-0047.

*Author Contributions:* AKMN performed the measurements and designed the setup; KB conceived the experiment; YP, BW and STP performed the molecular dynamics simulations; KB and AKMN wrote the manuscript and analyzed the data.